\documentclass[%
 reprint,
superscriptaddress,
 amsmath,amssymb,
 amsart,
 aps,
prl,
float
]{revtex4-1}

\usepackage{graphicx}% Include figure files
\usepackage{dcolumn}% Align table columns on decimal point
\usepackage{bm}% bold math
\usepackage{multirow}
\begin{document}

\preprint{APS/123-QED}

\title{High brightness CW electron beams from Superconducting RF photoemission gun}% Force line breaks with \\
%\thanks{}%

\author{I.~Petrushina}
%\email{ipetrushina@bnl.gov}
\affiliation{Department of Physics and Astronomy, Stony Brook University, Stony Brook, NY 11794, USA}
\affiliation{Collider-Accelerator Department, Brookhaven National Laboratory, Upton, NY 11973, USA}
\author{V.N.~Litvinenko}
\affiliation{Department of Physics and Astronomy, Stony Brook University, Stony Brook, NY 11794, USA}
\affiliation{Collider-Accelerator Department, Brookhaven National Laboratory, Upton, NY 11973, USA}
\author{Y.~Jing}
\affiliation{Department of Physics and Astronomy, Stony Brook University, Stony Brook, NY 11794, USA}
\affiliation{Collider-Accelerator Department, Brookhaven National Laboratory, Upton, NY 11973, USA}
\author{J.~Ma} 
\affiliation{Collider-Accelerator Department, Brookhaven National Laboratory, Upton, NY 11973, USA}
\author{I.~Pinayev} 
\affiliation{Collider-Accelerator Department, Brookhaven National Laboratory, Upton, NY 11973, USA}
\author{K.~Shih}
\affiliation{Department of Physics and Astronomy, Stony Brook University, Stony Brook, NY 11794, USA}
\author{G.~Wang}
\affiliation{Department of Physics and Astronomy, Stony Brook University, Stony Brook, NY 11794, USA}
\affiliation{Collider-Accelerator Department, Brookhaven National Laboratory, Upton, NY 11973, USA}
\author{Y.H.~Wu} 
\affiliation{Department of Physics and Astronomy, Stony Brook University, Stony Brook, NY 11794, USA}
\author{J.C.~Brutus}
\author{Z.~Altinbas}
\author{A.~Di Lieto}
\author{P.~Inacker}
\author{J. Jamilkowski}
\author{G.~Mahler}
\author{M.~Mapes}
\author{T.~Miller}
\author{G.~Narayan}
\author{M.~Paniccia}
\author{T.~Roser}
\author{F.~Severino}
\author{J.~Skaritka}
\author{L.~Smart}
\author{K.~Smith}
\author{V.~Soria}
\author{Y.~Than}
\author{J.~Tuozzolo} 
\author{E.~Wang}
\author{B.~Xiao}
\author{T.~Xin}
\affiliation{Collider-Accelerator Department, Brookhaven National Laboratory, Upton, NY 11973, USA}

\author{I.~Ben-Zvi}
\affiliation{Collider-Accelerator Department, Brookhaven National Laboratory, Upton, NY 11973, USA}
\author{C.~Boulware}
\affiliation{Niowave Inc., Lansing, MI 48906, USA}
\author{T.~Grimm}
\affiliation{Niowave Inc., Lansing, MI 48906, USA}
\author{K.~Mihara}
\affiliation{Department of Physics and Astronomy, Stony Brook University, Stony Brook, NY 11794, USA}
\author{D.~Kayran}
\affiliation{Department of Physics and Astronomy, Stony Brook University, Stony Brook, NY 11794, USA}
\affiliation{Collider-Accelerator Department, Brookhaven National Laboratory, Upton, NY 11973, USA}
\author{T.~Rao}
\affiliation{Collider-Accelerator Department, Brookhaven National Laboratory, Upton, NY 11973, USA}

\date{\today}% It is always \today, today,
             %  but any date may be explicitly specified

\begin{abstract}
CW photoinjectors operating at high accelerating gradients promise to revolutionize many areas of science and applications. They can establish the basis for a new generation of monochromatic X-ray free electron lasers, high brightness hadron beams, or a new generation of microchip production. In this letter we report on the record-performing superconducting RF electron gun with $\textrm{CsK}_{2}\textrm{Sb}$ photocathode. The gun is generating high charge electron bunches (up to 10 nC/bunch) and low transverse emittances, while operating for months with a single photocathode.  This achievement opens a new era in generating high-power beams with a very high average brightness.
%\begin{description}
%\item[Usage]
%Secondary publications and information retrieval purposes.
%\item[PACS numbers]
%May be entered using the \verb+\pacs{#1}+ command.
%\item[Structure]
%You may use the \texttt{description} environment to structure your abstract;
%use the optional argument of the \verb+\item+ command to give the category of each item. 
%\end{description}
\end{abstract}

\pacs{Valid PACS appear here}% PACS, the Physics and Astronomy
                             % Classification Scheme.
%\keywords{Suggested keywords}%Use showkeys class option if keyword
                              %display desired
\maketitle

%\tableofcontents

%\section{\label{sec:introduction}Introduction}

Superconducting radio-frequency (SRF) electron guns are frequently considered to be the favorite pathway for generating the high-quality, high-current beams needed for broad scientific and industrial applications, such as driving high-power X-ray and extreme ultraviolet (EUV) continuous wave (CW) free electron lasers (FELs) \cite{1,2,3,4,5,6,7,8,9,10}, intense $\gamma$-ray sources \cite{11, 12, 13, 14}, coolers for hadron beams \cite{15, 16, 17, 18}, and electron-hadron colliders \cite{19, 20, 21}. The quality of the generated electron beams---both the intensity and brightness---is extremely important for many of these applications. 

Next generation X-ray FELs (XFELs) are rapidly moving towards operation in CW mode due to the capabilities of providing more flexible photon pulse time patterns and higher average brightness.  This fact brings new requirements on the quality of the electron sources: they must demonstrate stable performance in CW mode, and deliver electron beams with low transverse normalized emittances \footnote{in this letter we will only quote the values of a normalized beam emittance $\varepsilon_{n}$.} (below 0.4~mm-mrad for 100~pC bunches \cite{aa}).

In this letter, we report the record performance of our SRF gun that was built for the Coherent electron Cooling (CeC) Proof of Principle (PoP) experiment at Relativistic Heavy Ion Collider (RHIC) \cite{15,16}. We will provide a brief overview of the current achievements in the performance of SRF guns worldwide, and show how the CeC photoinjector compares with these previously commissioned SRF guns. This is followed by the description of the emittance studies, including our experimental results and numerical simulations. We conclude this letter by demonstrating that our photoinjector meets the demands for the new generation of X-ray CW FELs.

Currently, there are three main types of photoinjectors: electrostatic (DC), pulsed RF, and CW RF electron guns. The DC guns, while generating superb CW beams \cite{29}, are limited in their maximum achievable accelerating gradients of 5~MV/m and kinetic energy below 0.5~MeV. In contrast, the pulsed normal conducting (NC) RF guns can reach accelerating gradients at the level of 100 MV/m, but operate at a relatively low repetition rate. CW RF guns can be based on either room temperature NC \cite{32,33} or SRF technology \cite{34,35,36,37,38,39,40,41,42,43}. The great potential of SRF guns was recognized as early as 1988 \cite{44}, and several successful experiments have been carried out since 2002 \cite{34,35,36,37,38}. A brief summary of the main experimental results in the performance of the five operational CW SRF photoinjectors is shown in Table~\ref{tab:est}.

%\section{{\label{sec:overview}Overview}}

\begin{table*}[bth!]
	\caption{\label{tab:est}The main experimental results for the five operational CW SRF photoinjectors.}
	\begin{ruledtabular}
		\begin{tabular}{lcccccc}
			Parameters & Units & CeC & HZDR \cite{34} & HZB \cite{42} & NPS \cite{37} & UW \cite{38}\\ \hline
			Cavity type &  -  & QWR \footnote{Quarter Wave Resonator} & Elliptical & Elliptical & QWR & QWR \\
			Number of cells & - & 1 & 3.5 & 1.4 & 1 & 1 \\
			RF frequency & MHz & 113 & 1300 & 1300 & 500 & 200 \\
			Operational temperature & K & 4 &	2 &	2 &	4 &	4 \\
			$E_{\textrm{max}}$ & MV/m & 18 & 12 & 7 & 6.5 & 12 \\ 
			
			Gun energy & MeV & 1.25 & 3.3 & 1.8 & 1.2 & 1.1 \\
			Bunch charge & nC & 10.7 & 0.3 & 0.077 & 0.078 & 0.1 \\
			Beam current & $\mu$A   & 140 & 18 & 0.005 & $<$0.0001 & $<$0.1 \\
			Dark current & nA & $<$1 & 30 & 100 & $<$20000 & $<$0.001  \\
			Photocathode & -  & $\textrm{CsK}_{2}\textrm{Sb}$ &	$\textrm{Cs}_{2}\textrm{Te}$ & Cu & Nb & Cu \\
			Laser wavelength & nm & 532 & 266 & 266 & 266 & 266\\
			\multirow{2}{*}{Projected emittance}& \multirow{2}{*}{mm-mrad} & 0.3 @ 100 pC & \multirow{2}{*}{2 @ 200 pC} & \multirow{2}{*}{0.5 @ 77 pC} & \multirow{2}{*}{4.9 @ 43 pC} & \multirow{2}{*}{1 @ 200 pC} \\
		    & & 0.57 @ 600 pC &  &  &  &  \\
		\end{tabular}
	\end{ruledtabular}
\end{table*}

%It is a well-known experimental fact that both, the 3D- and the projected, emittances\footnote{Normalized transverse emittance is defined as a physical emittance divided by $mc$ ($m$ is the electron's mass, and $c$ is the speed of light): $\varepsilon_{n} = \frac{1}{mc}\iint dxdP$. 3D normalized emittance is defined as $\varepsilon_{3Dn} = \frac{1}{(mc)^{3}}\iiint dxdP_{x}dydP_{y}dEdt$ where $x$, $y$ and $P_{x,y}$  are the transverse coordinates and Canonical momenta, $t$ is the arrival time of the particle, and $E$ is it’s energy.} grow with the increase of the bunch charge. 

To increase the attainable bunch charge and its brightness in a photoinjector, it is desirable to accelerate the electron beams to relativistic energies as quickly as possible, since we are operating in a space-charge dominated regime, and the space-charge effects are falling $\propto \dfrac{1}{\gamma^3}$ ($\gamma = \dfrac{E_{\textrm{tot}}}{mc^{2}}$, where $E_{\textrm{tot}}$ is the total energy of the beam, $m$ is the rest mass of the particles, and $c$ is the speed of light). However, there is always a limitation on the maximum density of the surface charge $\sigma$ that can be extracted from a photocathode \cite{22}:

\begin{equation}
    \sigma = \dfrac{E_{\textrm{em}}}{4\pi} ,
\end{equation}

\noindent
where $E_{\textrm{em}}$ is the electric field at the cathode at the moment of emission \footnote{we note that we are using Gaussian units in this letter}. Thus, the maximal bunch charge can be increased either by increasing the transverse size of the beam (and inevitably increasing its transverse emittance), or by increasing $E_{\textrm{em}}$.

%Second, a higher accelerating gradient is required for generating high-intensity and high-quality electron beams by accelerating them as quickly to relativistic energies\footnote{It is well known that space-charge forces are falling $\sim\dfrac{1}{\gamma^2}$, where  $\gamma = \dfrac{E}{mc^{2}}$.} \cite{22}. 

The overall dependence of the beam emittance from a photoinjector can be expressed as follows \cite{23,24,25,26,27,28,29,30,31}:

\begin{equation}
    \label{eq:square_root}
    \varepsilon_{n} \propto \sqrt{q\dfrac{E_{MTE}}{E_{\textrm{em}}}} ,
\end{equation}

\noindent
where $q$ is the bunch charge, and $E_{\textrm{MTE}}$ is the mean transverse energy of the photoelectrons at the cathode. The accelerating gradient at the moment of beam emission depends not only on the maximum electric field attainable in the RF gun, $E_{\textrm{max}}$, but also on the phase of emission $\varphi$, with $E_{\textrm{em}} = E_{\textrm{max}}\cdot\textrm{sin}\varphi$ \cite{22}. The phase is selected to maximize the beam energy gain. It depends on the geometry of the RF cavity, accelerating gradient $E_{\textrm{max}}$, the RF frequency $f_{\textrm{rf}} = \dfrac{\omega_{\textrm{rf}}}{2\pi}$, and is well described by a dimensionless parameter $\alpha$ \cite{28}:

\begin{equation}
    \alpha = \dfrac{eE_{\textrm{max}}}{2mc^{2}k_{\textrm{rf}}}
\end{equation}

\noindent
where $k_{\textrm{rf}} = \dfrac{\omega_{\textrm{rf}}}{c}$ is the RF wavenumber. The value of $\alpha$ indicates the relativism of the particle exiting the cavity [22,28], and determines the optimal phase of the emission: at $\alpha<1$, $\varphi$ is close to zero, and for $\alpha\gg 1$ it is close to the crest at $90^{\circ}$.

One can see that the choice of the geometry and operational frequency of a gun plays an important role. For example, the HZDR SRF gun \cite{34} operates at significantly higher frequency when compared to the CeC photoinjector. The maximum attainable electric field in the HZDR gun $E_{\textrm{max}}\sim$20 MV/m ($\alpha = 0.7$) results in the optimal emission phase of $10-15^{\circ}$. This fact limits the electric field at the cathode at the moment of emission $E_{\textrm{em}}\approx(0.2-0.25)\cdot E_{\textrm{max}}$. The CeC SRF gun operates at the same accelerating gradient, however the choice of the frequency results in $\alpha = 8.34$ with the optimal emission phase of $78.5^{\circ}$. This leads to a significantly higher electric field at the cathode during the emission $E_{\textrm{em}}\approx 0.98\cdot E_{\textrm{max}}$.

One major challenge for SRF photoinjectors is combining an RF cavity operating at cryogenic temperatures with photocathodes operating at room temperatures \footnote{Experiments have shown that operating $\textrm{CsK}_{2}\textrm{Sb}$ at the temperatures of liquid nitrogen \cite{39} reduces quantum efficiency (QE) by about one order of magnitude \cite{45}. Theory predicts that cooling it to liquid He temperatures will reduce the QE even further. In practice, replacing such a photocathode also is a long and complicated procedure.}. SRF guns with exchangeable photocathodes require an RF choke system, which provides an effective thermal insulation and ``grounds" the cathode to the nearby SRF cavity wall. The compatibility of the SRF environment of a cavity and the high quantum efficiency (QE) photocathodes was always questioned by the experts in the high-brightness electron sources \cite{cc}. The cathode or its insertion system can deposit lossy particulates on the walls of the cavity, which dramatically reduces its performance. On the other hand, the cavity can generate an ion and electron back-bombardment of the cathode, and cause its rapid degradation. The first hands-on experience with RF photoinjectors has proven that both of these challenges are very crucial for operation. This is why most of the SRF guns are operating with metal photocathodes \cite{37,38,42}. However, while being very robust, metal cathodes have a very low QE, and are not suitable for generating significant beam currents due to the heat load on the cathode by the laser power. The CeC SRF photoinjector utilizes room-temperature multialkali cathodes, which were specifically chosen to deliver high-quality electron beams. The first electron beam from the SRF photoelectron gun with 3~nC charge per bunch was generated in June 2015. It was a very important achievement indicating that our SRF gun and the $\textrm{CsK}_{2}\textrm{Sb}$ photocathodes are highly compatible.

\begin{figure}[htbp]
 	\includegraphics[width=1\linewidth]{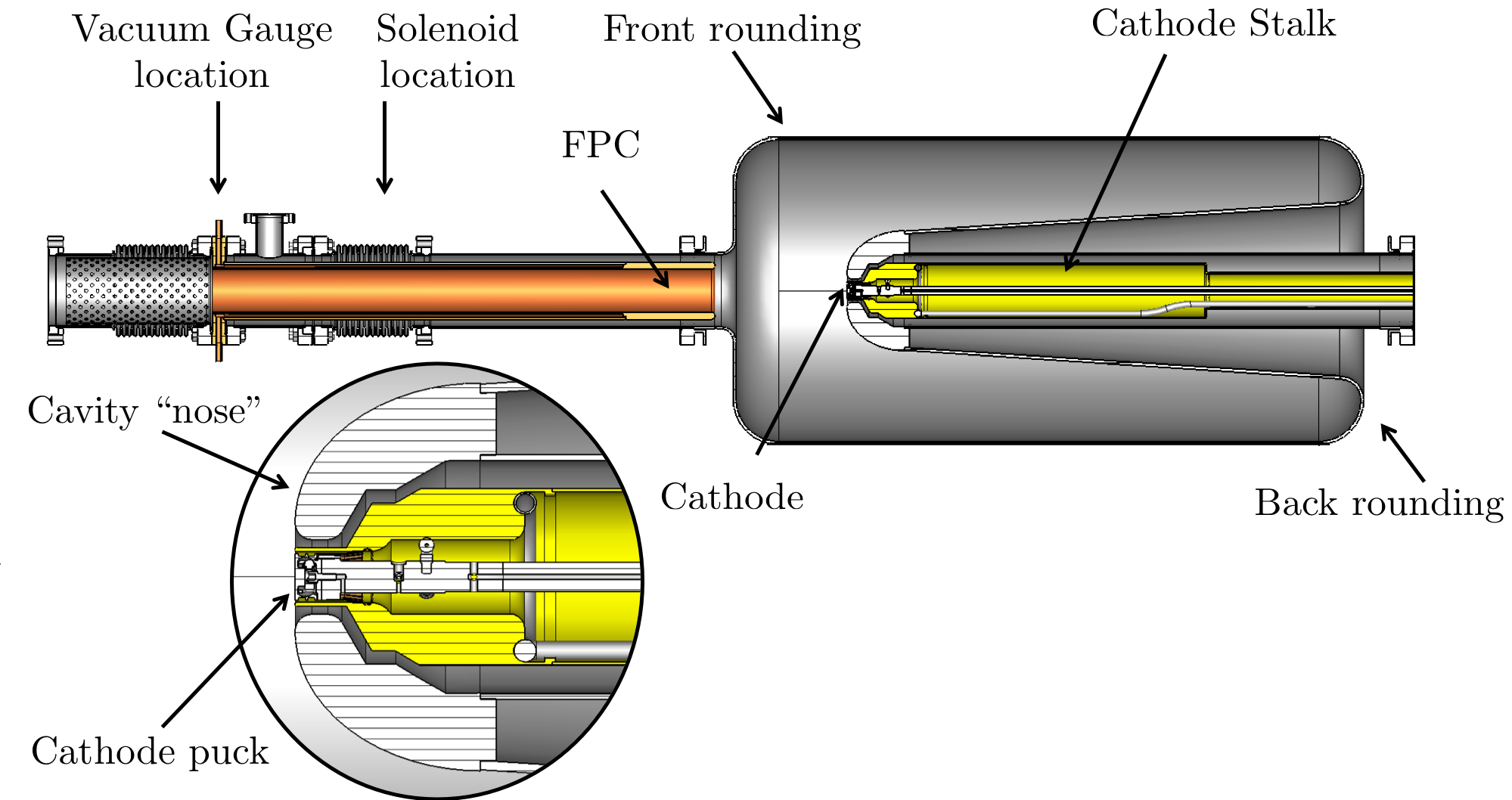}
 	\caption{Cross-section of the CeC PoP SRF gun. The inset demonstrates a detailed view of the cathode area.\label{fig:GunGeometry}}
 \end{figure}

The 113 MHz SRF cavity for the CeC PoP was the first of the designs based on a quarter wave resonator, built in 2010, which utilize low frequencies to produce long bunches with large bunch charge \footnote{The idea to use a quarter wave cavity as an electron source by accelerating along the symmetry axis was developed by Niowave and Brookhaven National Laboratory under the support of the DOE Small Business Innovation Research (SBIR) and Small Business Technology Transfer (STTR) program.}. Its detailed geometry and the main RF parameters can be found in Fig.~\ref{fig:GunGeometry} and Table~\ref{tab:Gun}, respectively.

%The 113~MHz SRF cavity for the CeC PoP experiment was built by Niowave Inc. in 2014, its detailed geometry and the main RF parameters can be found in Fig.~\ref{fig:GunGeometry} and Table~\ref{tab:Gun}, respectively.

Electron bunches---with typical duration of 400~ps---are generated from a $\textrm{CsK}_{2}\textrm{Sb}$ photocathode by 532~nm (green) laser pulses. The cathode is kept at room temperature inside the 113~MHz quarter-wave SRF Nb cavity operated at 4~K. 

The SRF gun accelerates a CW electron beam to a kinetic energy of 1.25 MeV (1.76 MeV total), and is operating at a repetition rate of 78~kHz, with 0.1-10.7~nC bunch charge.

The half-wavelength RF choke incorporates a hollow stainless steel cathode stalk which allows insertion of a cathode puck \cite{37,38}. The cathode stalk is gold plated to reduce heat emission into the 4 K system, and is kept at room temperature by circulating water. The stalk is shorted at the far end to serve as a choke filter. The axial position of the stalk tip and, therefore, the photocathode surface with respect to the cavity nose can be adjusted to optimize the focusing of the electron beam.   

\begin{table}[htbp]
	\caption{\label{tab:Gun}RF parameters of the gun.}
	\begin{ruledtabular}
		\begin{tabular}{lcc}
			Parameter & Units & Value\\ \hline
			Frequency & MHz & 113\\
			Quality Factor w/o Cathode & - & $3.5\times10^9$\\
			$R/Q$ & $\Omega$ & 126\\
			Geometry Factor& $\Omega$ & 38.2\\
			Operating Temperature& K & 4\\
			Accelerating Voltage& MV & 1.25-1.5\\
		\end{tabular}
	\end{ruledtabular}
\end{table}

The coaxial fundamental power coupler (FPC) with water cooling is incorporated in the front side of the cavity and allows for the beam to exit the gun. The FPC is placed on a motorized translation stage, so its position can be adjusted by 40~mm, which allows us to optimize the coupling and provide fine tuning of the gun frequency. The RF power to the gun is delivered from a 4-kW solid-state amplifier. 

\begin{table*}[htbp]
	\caption{\label{tab:EmittanceComparison}Comparison of the beam quality between the CeC SRF gun and the LCLS II injector requirements.}
	\begin{ruledtabular}
		\begin{tabular}{lccc}
		    Parameter & Units & LCLS II requirements & CeC SRF gun\\ \hline
			Gun voltage & MV & 0.75 & 1.5\footnote{Measured value}\\ 
			Charge per bunch & pC & 100 & 100-10,000\footnotemark[1]\\
			Average beam current @ 100 pC & mA & 0.062 & 0.15\footnotemark[1]\\
			Transverse RMS slice emittance @ 100 pC & mm-mrad & 0.4 & 0.15\footnote{Extracted from simulations} \\
			Transverse RMS projected emittance @ 100 pC & keV$\cdot$ps & - & 0.3\footnotemark[1] \\
			Longitudinal RMS slice emittance @ 100 pC & keV$\cdot$ps & 3.3 & 0.7\footnotemark[2] \\
			Quantum Efficiency & \% & 1 & 1-4\footnotemark[1] \\
		\end{tabular}
	\end{ruledtabular}
\end{table*}

The first focusing solenoid is located 65~cm downstream of the cathode surface, which is followed by the laser cross for the drive laser beam delivery. The CeC experiment is using a frequency doubled, in-house designed Nd:YAG MOPA (Master Oscillator Power Amplifier) system operating at 78~kHz. The system consists of a commercial, arbitrary waveform generator from iXblue and an in-house designed, solid state regenerative amplifier with a non-critical phase matching scheme for frequency doubling, producing 125-1000~ps, 100~$\mu$J pulses of 532~nm light.
The laser light is transported over 60~m and 3 optical tables connected with evacuated pipes to the electron gun, where a 1-10~mm variable aperture is illuminated and imaged onto the photocathode in a 1:1 ratio.

Our photoinjector allows for an in-situ cathode replacements using an ultra-high vacuum (UHV) manipulator system. A small (20~mm in diameter) cathode puck is made of molybdenum and has a high-QE $\textrm{CsK}_{2}\textrm{Sb}$ photoemission coating with diameter of 8 to 10~mm deposited in the center. The UHV portable transport system (which we call a ``garage") with built-in QE measuring system can store up to three cathode pucks without any significant loss of QE for a few months. Cathodes can be exchanged between the gun and the garage in about 30 minutes. The gold-plated RF spring finger contacts connect the puck with the inner surface of the stalk ensuring that the electromagnetic field does not propagate inside the cathode stalk.
 
% \begin{figure}[htbp]
%\begin{minipage}[t]{0.235\textwidth}
%\includegraphics[width=\textwidth]{Temp/CathodeNose.png}\\
%(a)
%\end{minipage}
%\begin{minipage}[t]{0.235\textwidth}
%\includegraphics[width=\textwidth]{Temp/CathodePuck.png}\\
%(b)
%\end{minipage}
%\caption{Section view of the stalk and the ``cavity nose" (a) and the cathode end assembly (b).}
%\label{fig:CathodePuck}
%\end{figure}

The electron beam from the gun can be evaluated using the intercepting diagnostics: two profile monitors and a Faraday cup for measuring the bunch charge. The Integrating Current Transformer (ICT) located at the exit of the gun system is used for a non-intercepting measurement of the electron beam current. 

%\section{\label{sec:performance}Performance}

%It took a significant effort to develop a reliable procedure for avoiding multipacting (MP) and operating the gun at the designed voltage. 
One of the challenges during the first year of the gun operation was the presence of strong multipacting (MP) which led to significant QE degradation of our cathodes. MP is a resonant process in which an electron avalanche builds up within a small region of a cavity surface resulting in an absorption of the RF power. Hence, MP can prevent the cavity from reaching the operational accelerating voltage and cause QE degradation. As the result of a thorough investigation \cite{46}, the repeatable and reliable procedure, which eliminated this challenge, was implemented. Throughout our 4-year long experience, we were capable of continuous operation of the high-QE (3-4\%) $\textrm{CsK}_{2}\textrm{Sb}$ photocathodes for months without any significant QE degradation.

%\begin{figure}[htbp]
%    \begin{minipage}[t]{0.49\linewidth}
%        \includegraphics[width=1\linewidth]{Temp/Charge_profile.pdf}\\
        
%        \centering
        
%        (a)
%    \end{minipage}
%    \begin{minipage}[t]{0.49\linewidth}
%        \includegraphics[width=1\linewidth]{Temp/QE_profile.pdf}

%        \centering
        
%        (b)
%    \end{minipage}
% 	\caption{QE map of one of the photocathodes after a month-long operation.\label{fig:QEmap}}
% \end{figure}

While we did not plan to establish any records, the CeC SRF gun has demonstrated an exceptional performance: in 2018 it generated electron bunches with charge exceeding 10~nC. 

%\section{\label{sec:emittance}Emittance Studies}

Since the primary focus of the CeC PoP experiment is not a characterization of the SRF gun performance, the available diagnostics for the emittance measurements is limited. We generally utilize a combination of a solenoid and a transverse beam profile monitor in order to evaluate performance of the photoinjector and measure the projected emittance. The measurements were compared with the simulations using ASTRA \cite{48}, GPT \cite{49} and PARMELA \cite{47} in order to achieve a better understanding of the delivered beam parameters.

\begin{figure}[h!]
 	\includegraphics[width=1\linewidth]{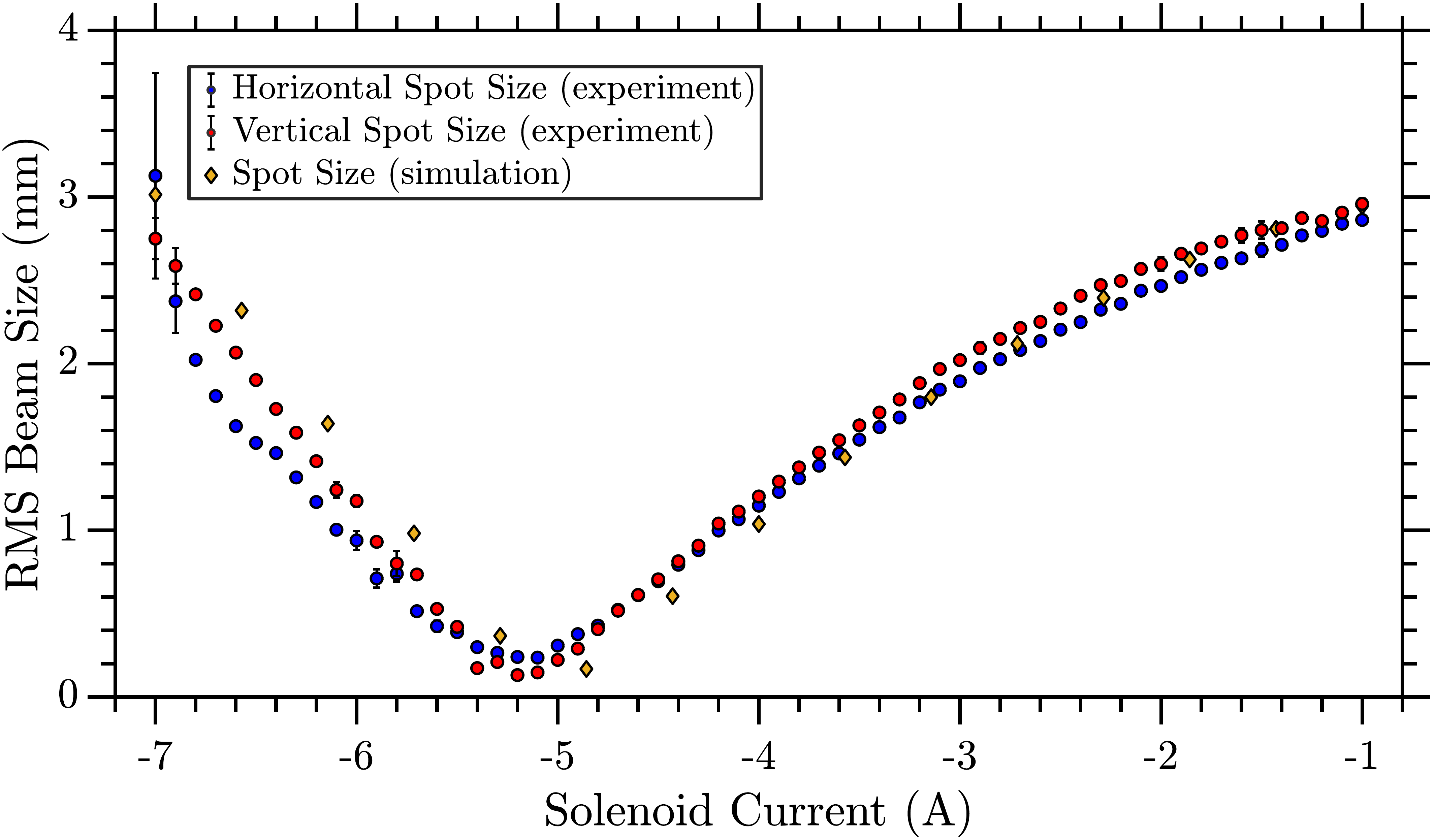}
 	\caption{Comparison of the solenoid scan for a 600~pC, 400~ps electron beam obtained through measurements (yellow diamonds) and simulations via ASTRA (red and blue circles). Corresponding projected and slice emittances are 0.57 and 0.35~mm-mrad, respectively.\label{fig:emittance}}
 \end{figure}

Figure~\ref{fig:emittance} demonstrates a comparison of the solenoid scan obtained experimentally and through simulations using ASTRA for a 600~pC, 400~ps electron beam. The results clearly indicate a good agreement between the measurements and numerical predictions. 

\begin{figure}[htbp]
 	\includegraphics[width=1\linewidth]{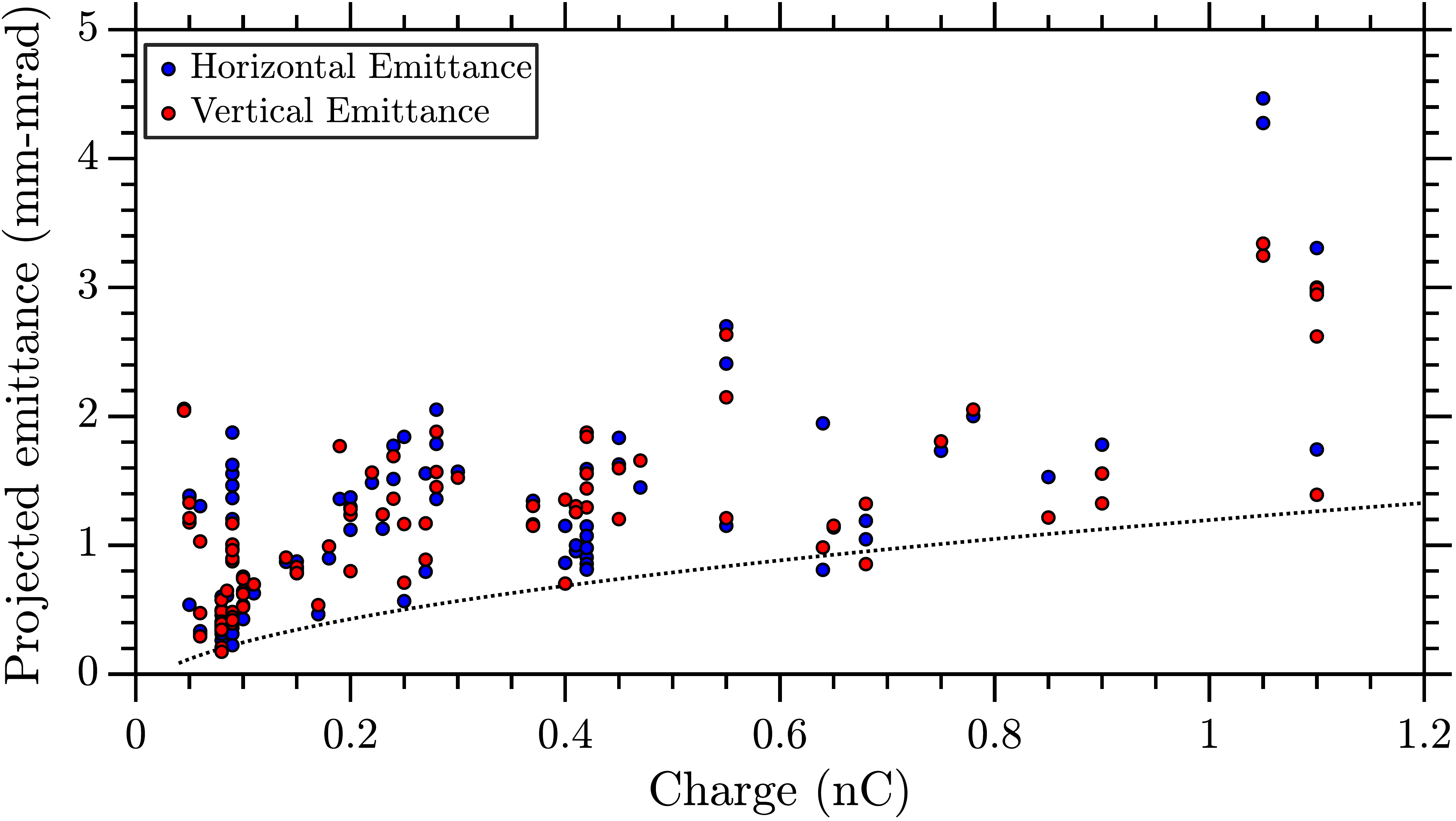}
 	\caption{Summary of the projected emittance measurements throughout the CeC operation in 2017-2018. Dashed curve is a fit for the lowest measured values of the emittance based on Eq.~\ref{eq:square_root}.\label{fig:emittance_meas}}
 \end{figure}

The summary of the projected emittance measurements performed during the CeC operation in 2017-2018 are shown in Fig.~\ref{fig:emittance_meas}. These data were obtained for a variety of conditions, and provide a good demonstration of the idea of emittance compensation---the measurements resulting in higher values of projected emittance show that the phase space ellipses of the beam slices were not properly aligned for those particular measurements.  

%The CeC PoP photoinjector has clearly demonstrated the best value of transverse emittance among the currently operational SRF guns, and even most of the operational NC guns. The experimental confirmation of the outstanding performance inspired us to present a possible setup based on our photoinjector which can be used as an electron source for modern X-ray FELs.

\begin{figure}[h!]
 	\includegraphics[width=1\linewidth]{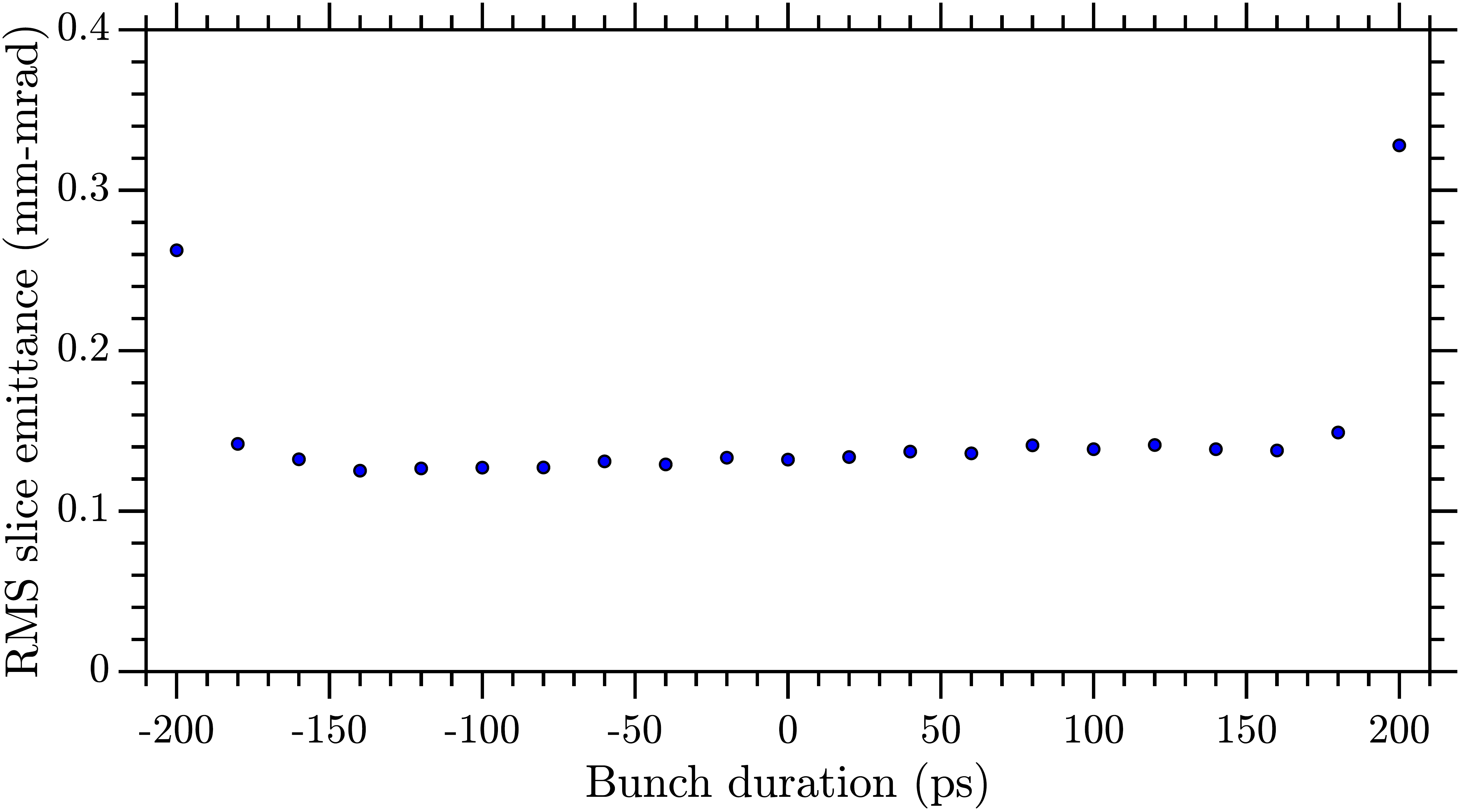}
 	\caption{RMS slice emittance obtained from the GPT simulation for a 100 pC, 400 ps electron beam.\label{fig:SliceEmittanceSimulation}}
 \end{figure}

During the experimental studies of our SRF gun it became apparent that it generates CW electron beams with quality satisfying the requirements for the CW X-ray FELs \cite{aa,bb}. Our SRF gun delivers CW electron beams with transverse and longitudinal emittances exceeding nominal requirements for the LCLS II CW X-Ray FEL \cite{aa}. A comparison of the beam quality for our SRF gun with the LCLS II injector requirements is shown in Table~\ref{tab:EmittanceComparison}. Figure~\ref{fig:SliceEmittanceSimulation} demonstrates that the CeC SRF gun can deliver 100 pC electron bunches with core slice emittance of 0.15~mm-mrad. Since the longitudinal emittance of the produced bunches is small, the desired peak current can be achieved via compression.

Hence, we demonstrated full compatibility of our SRF gun with the high-QE photocathodes, and the ability to deliver high-quality, high-brightness electron beams exceeding the requirements for the new generation of CW XFEL.

%\section{\label{sec:conclusion}Conclusion}

%\section*{Acknowlegments}
\begin{acknowledgments}

This research used resources of the National Energy Research Scientific Computing Center, which is supported by the Office of Science of the U.S. Department of Energy under Contract No.DE-AC02-05CH11231.

Work is supported by Brookhaven Science Associates, LLC under Contract No. DEAC0298CH10886 with the U.S. Department of Energy, DOE NP office grant DE-FOA- 0000632, and NSF grant PHY-1415252.

\end{acknowledgments}

\appendix
\label{appendix}

\bibliography{Petrushina_HighBrightness_CW_ebeams_from_SRF_gun_Bibliography}

\end{document}